\documentstyle[aps,prd,epsf]{revtex}

\def\bea{\begin{eqnarray}}
\def\eea{\end{eqnarray}}
\def\fr{\frac}

\def\r{\right}

\def\dag{\dagger}
\def\eps{\epsilon}

\def\eeas{\end{eqnarray*}}
\def\beas{\begin{eqnarray*}}
\def\ee{\end{equation}}
\def\be{\begin{equation}}

\def\iso{\vec\tau}

\def\veps{\varepsilon}

\def\dag{^\dagger}

\def\fpi2{\mbox{F$_\pi$}^2}

\def\mpi2{{m_\pi}^2}
\def\mk{m_K}
\def\mk2{{m_K}^2}

\def\fk2{\mbox{F$_K$}^2}
\def\df{F^\prime}

\def\r{\vec r}

\begin{document}

\draft

\title{\hfill {\rm  OUTP-99-29-P} \\ \hfill {\rm  TAN-FNT-99-04} \\
Multibaryons as Symmetric Multiskyrmions}

\author{
Juan P. Garrahan$^a$\thanks{E-mail: j.garrahan1@physics.ox.ac.uk.},
Mart\'\i n Schvellinger$^b$\thanks{E-mail: martin@venus.fisica.unlp.edu.ar}
and
Norberto N. Scoccola$^{c,d}$\thanks{Fellow of CONICET, Argentina. 
E-mail:scoccola@tandar.cnea.gov.ar}
}

\address{
$^a$ Theoretical Physics, University of Oxford,
1 Keble Road, Oxford, OX1 3NP, UK.\\
$^b$ Physics Department, University of La Plata,
C.C. 67, (1900) La Plata, Argentina.\\
$^c$ Physics Department, Comisi\'on Nacional de Energ\'{\i}a At\'omica,\\
	  Av.Libertador 8250, (1429) Buenos Aires, Argentina.\\
$^d$ Universidad Favaloro, Sol{\'{\i}}s 453, (1078) Buenos Aires, Argentina.}

\date{June 17, 1999}

\maketitle

\begin{abstract}
We study non-adiabatic corrections to multibaryon systems within the
bound state approach to the $SU(3)$ Skyrme model. We use approximate 
ans\"atze for the static background fields based on rational maps which have 
the same symmetries of the exact solutions. To determine the explicit form of 
the collective Hamiltonians and wave functions we only make use of these 
symmetries. Thus, the expressions obtained are also valid in the exact 
case. On the other hand, the inertia parameters and hyperfine splitting constants
we calculate do depend on the detailed form of the ans\"atze and are, therefore,
approximate. Using these values we compute the low lying spectra of multibaryons
with $B\le 9$ and strangeness $0$, $-1$ and $-B$. 
Finally, we show that the non-adiabatic
corrections do not affect the stability of the tetralambda and heptalambda
found in a previous work.    
\end{abstract}


\section{Introduction}

In the last few years there have been several important developments in the
determination of the lowest energy skyrmion configurations\cite{KS87,BTC90,BS97}. 
This type of solutions are essential for the understanding of multibaryons
and, perhaps, nuclei in the framework of the topological chiral soliton models. 
So far, these
models have proven to be useful for the description of quantities 
such as the masses, strong and electro-magnetic
properties of the octet and decuplet
baryons, baryon-baryon 
interactions, etc. (see e.g. Refs.\cite{ZB86,Wei96} and references therein). The 
knowledge of the properties of the multiskyrmion configurations opens the possibility 
of studying more complex baryonic objects. In fact, several investigations 
concerning non-strange multiskyrmion systems have been reported in the literature
(see, e.g., Refs.\cite{BC86,Car91,Wal91,Wal96,Irw98}).
Of particular interest are, however, the strange multibaryons. 
Perhaps the most celebrated example is the $H$ dibaryon
predicted in the context of the MIT bag model more than twenty years ago\cite{Jaf77}.
This exotic  
has been studied in various other models, including the Skyrme model
\cite{BBLRS84,KSS92,KM88,TSW94}, but its existence remains controversial both 
theoretically and experimentally.  
It has also been speculated that strange matter could 
be stable\cite{Wit84}. This has lead to numerous investigations of the properties of 
strange matter in bulk and in finite lumps (for a recent review see Ref.\cite{GSB98}). Moreover, 
with the new heavy ion colliders there is now the possibility of producing 
strange multibaryons in the laboratory\cite{E864}. In this situation the
study of multibaryon systems within the $SU(3)$ Skyrme model appears to be quite
interesting.  A first step in this direction has been reported in Ref.\cite{SS98}
where the rational map approximation\cite{HMS97} to the multiskymion fields was used
to describe the multibaryon configurations within the bound state approach\cite{CK85}
to the $SU(3)$ Skyrme model. Within this approach strange (multi)baryons appear
as systems of kaons bound to a background skyrmion configuration. To find the kaon 
binding energy one has to solve the corresponding eigenvalue problem. For a general
background this is a very hard numerical task since one has to deal with
several couple partial differential equations. However, this problem is greatly
simplified if one introduces the (approximate) rational maps ans\"atze for the 
multiskyrmion configurations. The construction of these ans\"atze is based
on the analogy between monopoles and skyrmions and requires that the
approximate solutions have the same symmetries than the exact numerical solutions.   
In fact, it is now known that up to $B=9$ these configurations are very
symmetric. Namely, for $B=2$ the solution corresponds to an axially 
symmetry torus while configurations with $B=3-9$ possess the symmetries
of the platonic polyhedra. In contrast with the exact solution,
however, the rational map approximation assumes that the modulus 
of the static pionic field is radially symmetric while its direction 
depends only on the polar coordinates. It was shown in Ref.\cite{HMS97} that
this represents a very good approximation. Once the rational
maps are introduced the kaon eigenvalue problem reduces, for each baryon number,
to one radial eigenvalue equation. The corresponding results have been given
in Ref.\cite{SS98}. In such reference, however, non-adiabatic effects were 
neglected. These effects appear when one performs the collective quantization
of the system. It should be stressed that it is only at this stage when the spin and 
isospin quantum numbers are well defined and splitting between the corresponding
states appears. The purpose of the present work is to carry out the collective
quantization of the bound multisoliton-kaon systems. This requires to pay special
attention to their symmetries which impose severe constraints on the possible
quantum numbers and wave functions.

This paper is organized as follows. In Sec. II we provide a brief description
of the model with special emphasis on the effect of the non-adiabatic corrections.
In Sec. III we describe in detail how to obtain the collective Hamiltonian for
the different baryon numbers, while in Sec. IV 
we focus on the corresponding wavefunctions. 
It should be noticed that since the discussions in these two sections rely only
on the symmetries of multiskyrmion configuration the corresponding
results hold true also for the exact solutions. In Sec. V we present the numerical
results and in Sec. VI our conclusions. Finally, in the Appendix
we give the explicit
form of the rational maps used in the present work.

\section{The Model}

We start with the effective action of the $SU(3)$ Skyrme model supplemented with an
appropriate symmetry breaking term\cite{Wei96}. Expressed in terms of the
$SU(3)$--valued chiral field $U(x)$ it reads
\bea
\Gamma = \int d^4x \ \left\{
{f^2_\pi \over{4}} Tr\left[ \partial_\mu U \partial^\mu U^\dagger \right] +
{1\over{32 e^2}} Tr\left[ [U^\dagger \partial_\mu U ,
U^\dagger \partial_\nu U ]^2 \right] \right\} + \Gamma_{WZ} +
\Gamma_{SB} \ ,  \label{action}
\eea
where $f_\pi$ is the pion decay constant (~$= 93 \ MeV$ empirically)
and $e$ is the so--called Skyrme parameter.  In Eq.(\ref{action}),
the symmetry breaking term $\Gamma_{SB}$ accounts for the different
masses and decay constants of the pion and kaon fields while
$\Gamma_{WZ}$ is the usual Wess--Zumino action. Their explicit
forms are
\bea
\Gamma_{SB}
& \! \! = \! \! &\int d^4x \left\{ {f_\pi^2  m_\pi^2 + 2 f_K^2 m_K^2 \over{12}}
Tr \left[ U + U^\dagger - 2 \right]
+ { f_\pi^2 m_\pi^2 - f_K^2 m_K^2 \over{6}}
Tr \left[ \sqrt{3} \lambda^8 \left( U + U^\dagger \right) \right] \right.
\nonumber \\
& & \hskip 1.cm \left. + { f_K^2 - f_\pi^2 \over{12} } Tr
\left[ \left( 1 - \sqrt{3} \lambda^8 \right)
\left( U \partial_\mu U^\dagger \partial^\mu U +
        U^\dagger \partial_\mu U \partial^\mu U^\dagger \right) \right]
\right\} \ ,
\label{oldmass}  \\
\Gamma_{WZ} \ &=& \ -i \fr{N_c}{240\pi^2}\int \ d^5x \
\veps^{\mu\nu\alpha\beta\gamma}
\ Tr(L_\mu L_\nu L_\alpha L_\beta L_\gamma) \ ,
\eea
where $\lambda^8$ is the eighth Gell-Mann matrix and $m_\pi$ and $m_K$
represent the pion and kaon masses, respectively,
and $f_K$ is the kaon decay
constant. 

We proceed by introducing the Callan--Klebanov ansatz for the
chiral field\cite{CK85}
\be
\label{ansatz}
U=\sqrt{U_\pi}U_K\sqrt{U_\pi} \ .
\ee
In this ansatz, $U_K$ is the field that carries
the strangeness. Its form is
\bea
U_K \ = \ \exp \left[ i\fr{\sqrt2}{{f_K}} \left( \begin{array}{cc}
							0 & K \\
							K\dag & 0
						   \end{array}
					   \right) \right] \ ,
\eea
where $K$ is the usual kaon isodoublet
$
K \ = \ \left( \begin{array}{c}
		   K^+ \\
		   K^0
		\end{array}
			   \right) \ 
$. The other component, $U_\pi$, is the soliton background field. It
is a direct extension to $SU(3)$ of the $SU(2)$ field, i.e.,
\bea
U_\pi \ = \ \left ( \begin{array}{cc}
    \exp\left[ \frac{i}{f_\pi} \vec \tau \cdot \hat \pi \right] & 0 \\
		       0 & 1
		    \end{array}
			       \right ) \ .
\eea

Replacing the ansatz Eq.(\ref{ansatz}) in the effective action
Eq.(\ref{action}) and expanding up to second order in the kaon
fields we obtain the Lagrangian density for the kaon--soliton system.
In the spirit of the bound state approach this coupled system
is solved by finding first the soliton background configuration. For 
this purpose we introduce the rational map ans\"atze\cite{HMS97}
\be
\vec \pi =  f_\pi \ \hat n \ F  \ ,
\label{mansatz}
\ee
with
\be
\hat n = {1\over{1+|R|^2}} \left( 2 \ \Re(R) \ \hat \imath +
				      2 \ \Im(R) \ \hat \jmath +
				     ( 1 - |R|^2 ) \ \hat k \right) \ ,
\label{pians}
\ee
where we have assumed that $F = F(r)$, and $R = R(z)$ is
the rational map corresponding to winding number $B=n$.
Here, $r$ is the usual spherical radial coordinate whereas the
complex variable $z$ is related to the other two spherical
coordinates $(\theta,\phi)$ via stereographic
projection, namely, $z = \tan(\theta/2) \exp(i \phi)$.
The resulting expression for the soliton mass per unit baryon is
(in what follows $s = \sin F$; $c = \cos F$)
\bea
M_{sol} = 
	{f_\pi^2\over{2n}} \int dr \ r^2 \left[ F'^2 + 2n\ {s^2\over{r^2}} 
        \left( 1 + {F'^2\over{e^2 f_\pi^2}} \right) + 
	{ {\cal I}\over{e^2 f_\pi^2}} 
       	{s^4\over{r^2}} + 8\pi m_\pi^2 ( 1 - c ) \right]
	\ .
\eea
The profile function $F(r)$ is obtained by minimizing $M_{sol}$ subject to the 
boundary conditions $F(0) = \pi$ and $F(\infty) = 0$. In using these 
boundary conditions we are assuming that all the extra winding number is obtained 
from the angular dependence of $\hat \pi$. The angular integral ${\cal I}$ is
\begin{equation}
{\cal I} \ = \ {r^4\over 16 \pi} \int d\Omega \ 
\left( \partial_i \hat n \cdot \partial_i \hat n \right)^2 \ = \ 
{1\over{4 \pi}} \int {2i\ dz d\bar z \over{ (1 + |z|^2)^2 }}
\ \left(
	 {{1 + |z|^2} \over{ 1 + |R|^2 }}
		  \left| {dR\over{dz}} \right| \right)^4 \ .
\end{equation}

In order to find the lowest soliton-kaon bound state we write the
kaon field as\cite{KM88,TSW94},
\be
\label{consatz}
K_{T_z}(\r,t) \ = \ k(r,t) \ \iso\cdot\hat n \ \chi_{T_z} \ ,
\ee
where $\chi$ is a two--component spinor.

The diagonalization of the
corresponding kaon Hamiltonian leads to the eigenvalue equation
\be
\left[ - {1\over{r^2}} \partial_r \left( r^2 h_n \partial_r \right)
+ m_K^2 + V^{eff}_n - f_n \epsilon_n^2	- 2 \ \lambda_n \ \epsilon_n
\right] k(r) = 0 \ .
\ee
Details on how to obtain this equation as well as  the explicit expression 
of the radial functions $f_n$, $h_n$, $\lambda_n$ and $V_n$ can be 
found in Ref.\cite{SS98}.

To obtain the hyperfine corrections to the multibaryons masses we
proceed with the semi-classical collective coordinates quantization
method, where the isospin and spatial rotations are treated as the
zero modes. Then, we introduce the time--dependent
spatial rotations $R$ and the isospin rotations $A$ such that
\bea
\label{rotpi}
u_\pi \ &\rightarrow & \ R \ A \ u_\pi \ A^{-1} \ , \\
\label{rotk}
K \ &\rightarrow & \ R \ A \ K \ .
\eea
The angular velocities with respect to the body fixed frame are given by
\bea
\label{angvel}
\left ( R^{-1} \dot R \right )_{ab} &=& \eps_{abc}\Omega_c \ , \\
A^{-1} \dot A &=& \fr{i}{2}\ \iso\cdot\vec \omega \ .
\eea
Replacing in the effective action we get the collective Lagrangian
\bea
L_{coll} = - M_{sol} + 
	{1\over2} \left[ \Theta^J_{ab} \ \Omega_a \Omega_b +
	\Theta^I_{ab} \ \omega_a \omega_b +
	2 \ \Theta^M_{ab} \ \Omega_a \omega_b \right]
	- \left( c^J_{ab} \Omega_a + c^I_{ab} \omega_a \right) \ T_b \ ,
	\label{lag}
\eea
where $a,b = 1,2,3$ and $T_b = (\chi^\dagger \tau_b \chi)/2$ is 
the kaon spin.

The moments of inertia $\Theta_{ab}$ and hyperfine splitting 
constants $c_{ab}$ appearing in 
Eq.(\ref{lag}) are given by
\bea
\Theta^J_{ab} &=& m_1 \ C_{ab} + \frac{m_2}{2} \ \bar C_{ab} \ , 
\label{tjab}
\\
\Theta^I_{ab} &=& m_1 \ (\delta_{ab} - A_{ab}) + 
         2 m_2 \ (n \ \delta_{ab} - \bar A_{ab} ) \ , 
\label{tiab} \\
\Theta^M_{ab} &=& m_1 \ B_{ab} + \frac{m_2}{2} \ \bar B_{ab} \ , \\
c^I_{ab} &=& \delta_{ab} - 3 \left[ ( \delta_{ab} - A_{ab}) \ d_1 + 
\frac{1}{2} \left( \bar A_{ab}+ 2 n A_{ab} \right) d_2 \right] \ , \\
c^J_{ab} &=& -3 \left[ B_{ab} \ d_1 + \left( \bar B_{ab} - n B_{ab}
\right) \ 
d_2 \right] \ , 
\eea
where the radial integrals $m_1$, $m_2$, $d_1$ and $d_2$ are 
\bea
m_1 &=& 4 \pi f_\pi^2 \int dr \ r^2 \ s^2 
\left( 1 + {F'^2 \over{e^2 f_\pi^2}} \right) \ ,  \\
m_2 &=& 4 \pi f_\pi^2 \int dr \ {s^4\over{e^2 f_\pi^2}} \ , \\
d_1 \ &=& \ 2\veps_n \ \int_0^\infty \ dr \ k^*k \ \left[ \fr{1}{3}r^2 f
( 1 + c ) \ -
\ \fr{1}{e^2\fk2}\fr{d}{dr} (r^2 \ \df s) \right ] \ , \\
d_2 \ &=& \ \fr{2\veps_n}{e^2\fk2} \ \int_0^\infty \ dr \ k^*k \;
\fr{2}{3} ( 1 + c) s^2 \ ,
\eea 
and the angular integrals
\bea
A_{ab} &=& \int {d\Omega\over{4\pi}} \ n^a n^b  \ , 
\label{inta} \\
\bar A_{ab} &=& r^2 \int {d\Omega\over{4\pi}} \ 
\partial_i \hat n\cdot \partial_i \hat n
\  n^a n^b \ ,  \\
B_{ab} &=& \int {d\Omega\over{4\pi}} \ \partial_b n^a \ ,   \\ 
\bar B_{ab} &=& r^2 \int {d\Omega\over{4\pi}} \ 
\partial_i \hat n\cdot \partial_i \hat n
\ \partial_b n^a \ , \\
C_{ab} &=& \int {d\Omega\over{4\pi}} \ 
\partial_a \hat n\cdot \partial_b \hat n \ , \\
\bar C_{ab} &=& r^2 \int {d\Omega\over{4\pi}} \ 
\partial_i \hat n\cdot \partial_i \hat n\ 
\partial_a \hat n\cdot \partial_b \hat n \ . 
\label{integrals}
\eea

The numerical values of these angular integrals depend 
only of the particular form of
the ansatz for $\hat n$ and not on the detailed 
form of the effective action and
its parameters. For the rational maps listed in the Appendix 
all the matrices Eqs. (\ref{inta})-(\ref{integrals}) are diagonal.
As we shall see in the next section, this is a direct consequence
of the symmetries of these ans\"atze. 
The corresponding values of the diagonal elements are listed in
Table I. Note that when all the diagonal elements are equal we list just one.
Also listed in Table I are the values of ${\cal I}$.
 
Given $L_{coll}$, the canonical momenta are then defined in the usual way
\begin{eqnarray}
\label{momenta}
J_a &=& \frac{\partial L_{coll}}{\partial \Omega_a} 
= \Theta_a^J \Omega_a + \Theta_a^M \omega_a - c_a^J \ T_a \ ,  \\
I_a &=& \frac{\partial L_{coll}}{\partial \omega_a} 
= \Theta_a^M \Omega_a + \Theta_a^I \omega_a - c_a^I \ T_a \ ,   
\end{eqnarray}
where we have used that, for the cases we are interested in, all 
the inertia and hyperfine splitting
constants are diagonal and thus denoted with a subindex $a=1,2,3$ the
corresponding diagonal elements. 
Depending on whether $\Delta_a \equiv \Theta_a^J \Theta_a^I - (\Theta^M_a)^2$
vanishes or not we have to follow a somewhat different procedure to obtain
the collective Hamiltonian. 
We consider first the case in which $\Delta_a \neq 0$ for
all values of $a$. In this case the relations Eq.(\ref{momenta})
can be inverted and the collective Hamiltonian results
\begin{equation}
H^{coll} = \sum_a \ H^{coll}_a \ , 
\label{nonvan}
\end{equation}
where
\begin{eqnarray}
H^{coll}_a &=& 
	\left( K_a^J  \ J_a^2 + K_a^I \ I_a^2 - 
	2 K_a^M \ J_a \ I_a \right) +
	2 \left( K_a^J \ \bar c_a^J \ J_a  + 
	K_a^I \ \bar c_a^I \ I_a \right) \ T_a \nonumber  \\
& & 
	+ \frac{K_a^I K_a^J}{K_a^I K_a^J-(K_a^M)^2} 
	\left( K_a^J \ (\bar c_a^J)^2 + 
	K_a^I \ (\bar c_a^I)^2 + 2 K^M_a \ \bar c_a^I \ \bar c_a^J
	\right) \ T_a^2
	\label{ha}
\end{eqnarray}
and
\begin{equation}
K_a^J =  \frac{1}{2} \frac{\Theta_a^I}{\Delta_a} \ ,  \quad
K_a^I =  \frac{1}{2} \frac{\Theta_a^J}{\Delta_a} \ ,  \quad
K_a^M =  \frac{1}{2} \frac{\Theta_a^M}{\Delta_a} \ ,  \quad  
\bar c_a^J = c_a^J - c^I_a \ \frac{\Theta^M_a}{\Theta^I_a} \ , \quad
\bar c_a^I = c_a^I - c^J_a \ \frac{\Theta^M_a}{\Theta^J_a} \ .
\end{equation}

If there exist, however, some values $i$ for 
which $\Delta_i = 0$ there appears a relation
between $I_i$, $J_i$ and $T_i$. It reads
\begin{equation}
J_i = \frac{\Theta_i^M}{\Theta_i^I} \ I_i - 
\left( c_i^J - c^I_i \ \frac{\Theta^M_i}{\Theta^I_i} \right) \ T_i \ .
\end{equation}
Using this relation it is not difficult to show that the collective 
Hamiltonian becomes
\begin{equation}
H^{coll} = \sum_{a\neq i} H^{coll}_a + 
\sum_i \frac{\left( I_i + c_i^I \ T_i \right)^2}{2 \Theta_i^I} \ .
\end{equation}
and the total multibaryon mass results
\begin{equation}
M = n \ M_{sol} + |S| \epsilon_n + E_{rot}
\end{equation}
where $S$ is the multibaryon strangeness and $E_{rot}$ the expectation
value of $H_{rot}$ in the corresponding wavefunction.
 
In the next section we will determine the precise form of the collective 
Hamiltonians for each baryon number.

\section{Collective Hamiltonians}

The minimum energy multiskyrmion configurations
are symmetric under certain groups of transformations\cite{BS97}. With 
the exception of the $B=1$ and $B=2$ cases where these symmetry groups are 
continuous ($O(3)$ and $D_{\infty h}$, 
respectively), these transformation groups have a finite number
of elements. 
In this section we will see how the symmetries of the multiskyrmion
configurations impose severe constraints on the detailed form of the 
collective Hamiltonian. For the $B \leq 4$
cases this has already been discussed in the literature using various
arguments. Here, we will extend such analysis within a unified
framework. It is important to notice that all the discussions and results
that follow are based only on the symmetries of the multiskyrmions.
Therefore, they will hold not only for the approximate configurations
based on the rational maps but also for the exact ones obtained from
numerical minimization.

The task here is to determine the precise structure of the
inertia and hyperfine splitting tensors, namely, which elements of
those tensors vanish and how many of the remaining non-zero elements
are independent for each baryon number. First, we note that
each operation of the abstract group $G$ is represented by a pair 
of operations $\{g, D_g\}$ which act in spin and isospin spaces, 
respectively. The pion field in Eq.(\ref{pians}) is invariant under 
these combined operations, 
\begin{equation}
\vec \tau \cdot \vec \pi (\hat r) = D_g  \vec \tau \cdot \vec \pi 
(g^{-1} \hat r) (D_g)^\dagger \ .
	\label{dege}
\end{equation}
Given the form used for the kaon field\footnote{This ansatz
can be easily generalized if the exact numerical soliton configuration
is used instead of the approximation based on rational maps.}, 
Eq.(\ref{mansatz}), this invariance implies that the action 
of the group element on the kaon field is also 
represented by $D_g$. In fact,
\begin{equation}
D_g K_{T_z} (\vec r, t) \ =  K_{T_z} (g \vec r,t) \ ,
	\label{tege}
\end{equation}
which means that the symmetry operation acting on the 
kaon field is just given 
by the representation of the isospin operation $D_g$ in the $T$-space.
Thus, the $\vec I$ and $\vec T$ operators transform in the
same way under elements of $G$. This shows that it 
is enough to perform the 
explicit analysis only for the inertia tensors. 
Once this is done the results 
for the hyperfine splitting constants can be easily obtained noting that
in Eq.(\ref{ha}),  
$c^J_{ab}$ plays a role similar to that of $K^M_{ab}$, 
while $c^I_{ab}$ to that of $K^I_{ab}$. 

The inertia tensors can be diagonalized by an appropriate choice of the
spatial and internal reference frames, and this is in fact what
happens for the rational map ans\"atze given in the Appendix.
Consider first the case for the spin. 
The spin generators $J_a$ transform under $G$ in some
(possibly reducible) representation. 
The number of independent diagonal components of
the inertia tensor (moments of inertia) will be equal to the number of
irreducible representations\footnote{The 
character tables containing the list of irreps of the groups
we are interested in can be found, e.g., in
Refs.\cite{Wil55} and \cite{KDWS63}. 
We follow the conventions of Ref.\cite{Wil55}.} (irreps)
of $G$ into which this representation breaks,
since the combination $K^J_{ab} \, J_a J_b$ must be a scalar
under $G$. 
The spin generators belong to the 
$1^+$ irrep of $O(3)$ which for the cases we will consider below
breaks into 
either a 3-dim irrep or as the sum
of 1- and 2-dim irreps of $G$. 
In the first case there is only one
moment of inertia and the spin Hamiltonian is proportional to 
$\sum_a J_a J_a$, while in the second case there are two moments, and
the
Hamiltonian contains the terms $J_1^2+J_1^2$ and $J_3^2$. 
The same argument holds for the other collective operators.

An important remark is the following.
While there is a one-to-one correspondence 
between $g$ and the elements of $G$, 
this is not necessarily the case for the operations $D_g$. 
In other words, it 
could happen that the same $D_g$ is associated with two (or more) 
different elements 
in spin space. 
In this case, the operations $D_g$ do not span the full 
group $G$ but a subgroup 
of it. As a consequence, the generators $J_a$ and $I_a$ ($T_a$)
could transform in different
representations of $G$. This would imply that the corresponding 
mixing inertia
would vanish. Below we see that this happens for some values of $B$.

Let us consider now the multiskyrmion configurations case by case. 
The $B=1$ skyrmion is spherically symmetric\cite{ZB86}. Thus, 
the relevant symmetry group $G$ is $O(3)$. 
In this case, $g=D_g$ and 
both $\vec J$ and $\vec I$ are in the 3-dim irrep $1^+$. Using the 
arguments given above we have
\begin{equation}
\Theta_{a}^J = \Theta^J \ , \qquad    
\Theta_{a}^I = \Theta^I \ , \qquad    
\Theta_{a}^M = \Theta^M \ , \qquad 
c_{a}^J = c^J \ , \qquad 
c_{a}^I = c^I \ .
\end{equation}
Since in this case we are dealing with a continuous group the equality between
the representation of the group elements in spin and isospin spaces can
be written in terms of corresponding generators of the algebra. Namely,
we obtain the relation $J_a = I_a + T_a$. From Eq.(\ref{momenta}) this
implies
\begin{equation}
\Theta^J = \Theta^I = \Theta^M \ , \qquad c^I = 1 - c^J \ ,
\end{equation}
which leads to $\Delta_a=0$ for all values of $a$. 
Then, the collective Hamiltonian
takes the well-known form 
\begin{equation}
H^{coll}_{B=1} = \frac{1}{2 \Theta} 
\left( I^2 + c^2\ T^2 + 2 \ c \ \vec T \cdot \vec I \right) \ .
\end{equation}

As already mentioned, 
the $B=2$ lowest energy skyrmion configuration has the symmetry of a 
torus\cite{KS87} which implies $G = D_{\infty h}$. Choosing the symmetry axis
along the z-direction we obtain that the third 
components of the momenta are 
in the 1-dim $\Sigma^-_g$ while the other 
two components are in the 2-dim irrep 
$\Pi_g$.
Since rotations along the z-axis form a continuous 
subgroup of  $D_{\infty h}$ we obtain
for the terms containing third components of the 
momenta a result similar to that of
$B=1$,
\begin{equation}
\Theta^J_3 = \Theta^I_3 = \Theta^M_3  \ , 
\qquad c^I_3 = 1 - c^J_3  \ , 
\end{equation}
which leads to $\Delta_3 = 0$. For the other 
components $\Delta_{1,2} \neq 0$ since
the $C_2$ along those axes only form finite subgroups of $G$.  
Consequently, the corresponding component of the 
different type of inertia
and splitting constants need not to be equal and the 
$B=2$ collective Hamiltonian reads
\begin{eqnarray}
H^{coll}_{B=2} &=& 
	K_1^J \left(J^2 - J_3^2\right) + K_1^I \left(I^2 - I_3^2\right) 
	+ K_1^I (\bar c_1^I)^2 \left(T^2 - T_3^2\right) \nonumber \\
	& & + K_1^I \bar c_1^I \left(I_+ \ T_- + I_- \ T_+ \right) + 
    	\frac{(I_3 + c_3^I T_3)^2}{2 \Theta_3^I} \ .
\label{hdos}
\end{eqnarray}

For the rest of the baryon numbers under consideration, 
$B=3-9$, the symmetry group
$G$ is finite\cite{BTC90,BS97}. Therefore, $\Delta_a$ never vanishes 
for all those baryon numbers
and the collective Hamiltonian will have the general 
form Eq.(\ref{nonvan}). There
can be, however, some further simplifications depending 
on the way in which the symmetry 
is realized in spin and isospin spaces. 
   
The symmetry group of the $B=3$ solution is $G=T_d$. In this case, we have 
that $g=D_g$ for all the elements of $G$\cite{Car91}. Thus,
the components $J_a$, $I_a$ and $T_a$ are in the 3-dim irrep $F_2$. 
The collective Hamiltonian reads
\begin{eqnarray}
H^{coll}_{B=3} &=& K^J \ J^2 + K^I \ I^2 - 2 K^M \ \vec I \cdot \vec J + 
2 K^J \bar c^J 
\vec J \cdot \vec T + 2 K^I \bar c^I 
\vec I \cdot \vec T \nonumber \\
& & + \frac{ K^I K^J }{ K^I K^J - (K^M)^2 } 
\left(K^J \ (\bar c^J)^2 + K^I \ (\bar c^I)^2 + 2 K^M \ \bar c^I \ \bar c^J
\right) \ T^2 \ .
\label{btres}
\end{eqnarray}

In the case of $B=4$ the relevant symmetry group is $O_h$. As discussed
in Ref.\cite{BTC90}, for the minimum energy configuration this symmetry
is realized in such a way that the elements $D_g$ cover four times
the $D_{3d}$ subgroup. As a result, $I_1$ ($T_1$) and $I_2$ ($T_2$) are in the 2-dim irrep
$E_g$, $I_3$ ($T_3$) in the $A_{2g}$ irrep and 
the components of $\vec J$ lie in the 3-dim irrep $T_{1g}$. 
We see then that the mixing inertia and spin
splitting tensors vanish. The resulting form of the
corresponding collective Hamiltonian is  
\begin{eqnarray}
H^{coll}_{B=4} &=& K^J \ J^2 + K_1^I \ ( \vec I + \bar c_1^I \vec T )^2  
+ (K_3^I - K_1^I) \ I_3^2 \nonumber \\
& & + 2 ( K_3^I \bar c_3^I - K^I_1 \bar c_1^I ) \ I_3 \ T_3 + 
(K^I_3 (\bar c^I_3)^2 - K^I_1 (\bar c^I_1)^2 ) \ T_3^2 \ .
\end{eqnarray}

The lowest energy multiskyrmion with $B=5$ has
$D_{2d}$ symmetry. In this case, there is a one-to-one correspondence
between the realization of the group in spin and isospin spaces.
It is easy to check that the third components of the momenta
are in the $A_2$ irrep while the other two components in the
2-dim one $E$. The resulting collective Hamiltonian is
\begin{eqnarray}
H^{coll}_{B=5} &=& K_1^J \ ( J^2 - J_3^2 ) + K_1^I \ ( I^2 - I_3^2 ) - 
2 K_1^M \ ( \vec I \cdot \vec J - I_3 \ J_3 ) 
\nonumber \\
& & + 
2 K_1^J \bar c_1^J ( \vec J \cdot \vec T - J_3 \ T_3 ) +
2 K_1^I \bar c_1^I ( \vec I \cdot \vec T - I_3 \ T_3 ) 
\nonumber \\
& & + K_3^J \ J_3^2 + K_3^I \ I^2_3 - 2 K_3^M  \ I_3 \ J_3 + 
2 K_3^J \bar c_3^J \ J_3 \ T_3 + 
    2 K_3^2 \bar c_3^I \ I_3 \ T_3 
\nonumber \\
& & + \frac{ K_1^I K_1^J }{ K_1^I K_1^J - (K_1^M)^2 } 
\left(K_1^J \ (\bar c_1^J)^2 + K_1^I \ (\bar c_1^I)^2 + 
2 K_1^M \ \bar c_1^I \ \bar c_1^J
\right) \ ( T^2 - T^2_3 ) 
\nonumber \\
& & +  
\frac{K_3^I K_3^J}{K_3^I K_3^J - (K_3^M)^2} \left( K_3^J (\bar
c_3^J)^2 + 
K_3^I (\bar c_3^I)^2 
+ 2 K_3^M \bar c_3^I \bar c_3^J \right) T_3^2 \ .
\end{eqnarray}

As found in Ref.\cite{BS97}, for $B=6$ the symmetry group is $D_{4d}$.
Due to the way in which the generators of the group are realized as
pairs of spin-isospin operations it is
possible to show that while the spin operations cover the full
$D_{4d}$ group the isospin one cover twice the $D_{2d}$ subgroup.
From the corresponding compatibility tables together with the 
compatibility table of the full rotational group we find that
$J_3, I_3$ and $T_3$ transform as the $A_2$ irrep,  $J_1$ and $J_2$ as
the $E_3$ irrep and the rest as $E_2$ irrep. Therefore, 
\begin{eqnarray}
H^{coll}_{B=6} &=& K_1^J \ J^2 +  K_1^I \ (\vec I + \bar c_1^I \vec T)^2 
+ (K_3^J - K_1^J) \ J_3^2 + (K_3^I - K_1^I) \ I^2_3 
\nonumber \\
& & - 2 K_3^M \ I_3 \ J_3 + 2 K_3^J \bar c_3^J \ J_3 \ T_3 + 
2 \left( K_3^I \bar c_3^I 
- K_1^I \bar c_1^I \right) \ I_3 \ T_3 \nonumber \\
& & + \left[ \frac{K_3^I K_3^J}{K_3^I K_3^J - (K_3^M)^2} \left( K_3^J
(\bar c_3^J)^2 + 
K_3^I (\bar c_3^I)^2 
+ 2 K_3^M \bar c_3^I \bar c_3^J \right) - K_1^I (\bar c_1^I)^2 \right] T_3^2 
\label{bseis} \ .
\end{eqnarray}

The $B=7$ configuration has icosahedral symmetry $I_h$ with the symmetry
realized in such a way that the components of the spin operators transform 
like the $F_{1g}$ irrep while those of the isospin operators as $F_{2g}$
irrep. Thus, the collective Hamiltonian takes the simple form  
\begin{equation}
H^{coll}_{B=7} = K^J \ J^2 + K^I \ ( \vec I + \bar c^I \vec T)^2 \ .
\end{equation}

For $B=8$ we have to deal with the $D_{6d}$ group. Like the case
of lower even baryon numbers the isospin 
operations do not span the full group but twice a subgroup, $D_{3d}$
in this case. We find that $J_3, I_3$ and $T_3$ transform as the 
$A_2$ irrep, $J_1$ and $J_2$ as $E_5$ irrep and the rest as the $E_4$
irrep. This implies that the collective Hamiltonian for $B=8$ has the same 
form as the $B=6$ one given in Eq.(\ref{bseis}).  
Finally, the $B=9$ multiskyrmion has the same symmetry as the $B=3$ one,
$T_d$. Consequently, we obtain a similar form for the corresponding
collective Hamiltonian, Eq.(\ref{btres}).

\section{Collective Wave Functions}

Having determined the explicit form of the collective Hamiltonian
we have to find the corresponding wave functions. These
wavefunctions have to satisfy some constraints imposed by the symmetries
of the background multiskyrmion. For non-strange multiskyrmions this
problem has been discussed by several authors\cite{BC86,Car91,Wal91,Wal96,Irw98}. 
Here, we will extend such studies for kaon-soliton bound systems.

The quantization of a single Skyrmion as a fermion implies that 
under certain symmetry operations of the classical multisoliton background the corresponding 
wave functions can pick up 
a nontrivial phase. These are known as Finkelstein-Rubinstein (FR) constraints
\cite{firu}.
We can generically write the constraints 
on the ground state as 
\begin{equation}
g \ D_g | \mbox{g.s.} \rangle = \gamma_g | \mbox{g.s.} \rangle \ , 
\label{cons}
\end{equation}
where $\gamma_g = \pm 1$ is determined according to the FR constraints.
Using
continuity arguments it turns out that the FR phases can be non-trivial
only for those operations corresponding to rotations, so for our cases 
of interest only the proper subgroup of $G$ needs
to be considered. For the isospin transformations we have to 
take into account 
the fact that the symmetry operation also acts on the kaon field.
From Eq.(\ref{tege}), however, we notice that this operation coincides
with the one acting on the soliton isospin space.
Thus, defining $\vec N = \vec I + \vec T$, the problem basically
reduces to that of non-strange baryons just replacing the collective isospin 
by $\vec N$. The (proper) group generators and their corresponding FR phases for 
the configurations considered in this work were determined in 
Refs.\cite{Car91,Irw98}. They are listed in Table \ref{symop}.

It is clear from Eq.(\ref{cons}) that due to the FR phases the 
soliton ground state might transform in a one-dimensional  
non-trivial irrep of $G$. 
Using the FR phases
listed in Table \ref{symop} and the group character tables, the 
relevant 1-dim irrep $\Gamma$ can be determined. We obtain that, except for the
$B=5$ and $B=6$ cases, all the wavefunctions should transform as the 
trivial irrep of the corresponding symmetry groups. For $B=5$,
$\Gamma$ is the $A_2$ irrep of $D_{2d}$ while for
$B=6$ the wave functions should transform as the $A_2$ irrep
of $D_{4d}$.

We now need to determine the collective wavefunctions. The general
procedure for arbitrary soliton backgrounds was discussed
in \cite{GK97}.
We consider first the problem without strangeness. In this case we
need to determine the functions
\begin{equation}
| J J_z, I I_z \rangle = \sum_{J_3 I_3} \,
	\alpha^{JI}_{J_3 I_3} \, 
	D^J_{J_z J_3} \, D^I_{I_z I_3} \ ,
	\label{jii}
\end{equation}
which transform under the {\em right} action of $G$ in the irrep
$\Gamma$ of the soliton. This can be done following standard group
theoretical methods \cite{S93}.
The product representation $J \times I$
of $SU(2)$ is in general a reducible representation of $G$. 
The projector operator into the irrep $\Gamma$ is
\begin{equation}
P_{\Gamma} = \frac{1}{|G|} \sum_{g \in G} \,
	\chi_{\Gamma}^*(g) \, \rho(g) \ ,
	\label{proj}
\end{equation}
where $|G|$ is the rank of the group, $\chi_{\Gamma}(g)$ the
character of operation $g$, and $\rho(g)$ the representation of $g$
in $J \times I$  (cf. Eq.(\ref{dege}))
\begin{equation}
\rho(g) = D^J(g) \times D^I(D_g) \ . 
\end{equation}
The eigenvalues of $P_{\Gamma}$ can either vanish or be equal
to one. The eigenvectors corresponding to each non-vanishing 
eigenvalue provide precisely the coefficients $\alpha^{JI}_{J_3 I_3}$
of Eq.(\ref{jii}), and there are as many wave functions as
non-zero eigenvalues. If all eigenvalues vanish there is no collective
state with the given $J,I$. If there is only one, the wavefunction is
an eigenfunction of the collective Hamiltonian, and if there are more
than
one, the Hamiltonian has to be diagonalized in the subspace
spanned by them.

Let us proceed now to the case with $S \neq 0$. We need to find the
functions\footnote{Note that $T = |S|/2$. See below.} 
\begin{equation}
| J J_z, I I_z, S \rangle = \sum_{J_3 I_3 T_3} \,
	\beta^{JIT}_{J_3 I_3 T_3} \, 
	D^J_{J_z J_3} \, D^I_{I_z I_3} \, K^T_{T_3} \ ,
  	\label{jit}
\end{equation}
which transform in irrep $\Gamma$ under $G$. However, as noted above,
the action of $G$ in isospin and $T$-spaces is the same, so it is
possible to couple them to $\vec N = \vec I + \vec T$.
Our problem then reduces to that of the case without strangeness:
for given $I$ and $S$ we have several possible values of $N$,
for each of these we determine the linear combinations Eq. (\ref{jii})
with $I$ replaced by $N$, and finally we uncouple $I$ and $T$. 
We obtain
\begin{equation}
| J J_z, I I_z, S \rangle =
	\sum_{J_3 N_3 I_3 T_3} \,
	\alpha^{JN}_{J_3 N_3} \,
	\langle I I_3 T T_3 | N N_3 \rangle \,
	D^J_{J_z J_3} \, D^I_{I_z I_3} \, K^T_{T_3}
\end{equation}
where $\langle I I_3 T T_3 | N N_3 \rangle$
are the $SU(2)$ Clebsch-Gordan coefficients. 

There is a further restriction of the possible collective states. 
Given a certain value of the baryon number $B$ and
the strangeness $S$, not all the values of isospin $I$ are allowed. 
As discussed in Appendix
B of Ref.\cite{TSW94}, physical states should have hypercharge 
and isospin given by
\begin{equation}
Y = B + S/3 = {{p + 2 q}\over 3} \qquad  ;   
\qquad  
I = {p\over2}
\end{equation}
where $p$ and $q$ should be non-negative integer numbers. 
The allowed values of isospin 
$I$ for
states with $S=0,-1$ and $-B$ are given in Table \ref{isospin},
together with the corresponding
values of $T$. Such values are obtained by imposing that the 
kaon wave function has to be
completely symmetric under individual kaon exchange. 

It should also be noted
that in the construction of the projector Eq. (\ref{proj})
all the operations of $G$ have to
be taken into account (i.e. not only those of the proper subgroup). 
For this purpose the
representations of the parity operation are also needed. 
For each baryon number, they 
are given in Table \ref{symop}. 
Another important comment is that for odd baryon 
numbers the $J$ and $N$ quantum numbers are half-integers. 
For those cases
one has to deal with the double group of $G$.

\section{Numerical results}

In our numerical calculations we will use two
standard sets of values for the Skyrme model parameters
$f_\pi$, $e$ and $m_\pi$.  
SET A corresponds to $f_\pi = 64.5 \ MeV$, $e=5.45$, $m_\pi = 0$ while SET B to  
$f_\pi = 54 \ MeV$, $e=4.84$, $m_\pi = 138 MeV$ \cite{ANW83}.
In both cases we set
the ratio $f_K/f_\pi$ to its empirical ratio $f_K/f_\pi = 1.22$.
With these values we can calculate $M_{sol}$, the kaon eigenenergies $\epsilon_n$
and the radial integrals $m_1$, $m_2$, $d_1$ and $d_2$ which appear
in the expression of the moments of inertia and hyperfine splitting constants. 
The results are tabulated in Table IV. Using these values together with those for 
the angular integrals given in Table I all the parameters appearing in the collective
Hamiltonians can be evaluated. For $B=1$ we find that $\Theta = 1.01 \ fm$ and 
$c=0.50$ for Set A and  $\Theta = 1.01 \ fm$ and $c=0.39$ for Set B which provide
a quite accurate description of the octet and decuplet baryon 
spectra\cite{CK85,RS91}.  The numerical values of the parameters in the
$B=2$ collective Hamiltonian Eq.(\ref{hdos}) are given in Table V. It interesting
to compare the values of the inertia parameters with those obtained using
the numerically obtained exact axially symmetric $B=2$ skyrmion\cite{KS87}. For example,
the corresponding values for Set B are
\begin{equation}
K_1^J = 30 \ MeV \ , \quad K_1^I = 48 \ MeV \ , 
	\quad \Theta_3^I = 1.45 \ fm \ .
\end{equation}
As we see the differences with the values listed in Table V are of only a few percent. 
On the other hand 
there not exist, so far, any
calculation of hyperfine splitting constants using the exact numerical $B=2$ skyrmion.
Nevertheless, we can compare our results with those from a calculation based on an
improved variational ansatz\cite{TSW94} which are, for Set B, 
\begin{equation}
\bar c_1^I = 0.334  \ ,  \qquad c_3^I = 0.554 \ .
\end{equation} 
These values are also very similar to ours. 
This is also true for Set A. Taking into account that the corresponding 
inertia parameters are also very close to those given in Table V, it follows that our 
predicted dibaryon spectra coincide basically with the ones described in Ref.\cite{TSW94}. 

Results for the $B=3-9$ inertia parameters and hyperfine splitting 
constants are listed in Tables VI and VII, respectively. As expected,
the inertia parameters decrease with increasing baryon number. 
However, the decrease of the spin inertia appears to be much faster than that of 
the isospin one. This can be understood in the following way. Since we are interested
in the overall behavior of inertias as a function of $B$ we define, for both
spin and isospin, the average value $K = 1/3 \sum_a K_{aa}$. 
As it can be seen from Table IV, 
while $m_1$ is roughly proportional to the baryon number, $m_2$ is basically
independent of $B$. Therefore, assuming $K \approx 1/\Theta$ and 
using Eqs.(\ref{tjab},
\ref{tiab}) we have
\begin{equation}
1/K^J \approx a \  n \  Tr C + b \  Tr \bar C  \ , \qquad
1/K^I \approx a \  n \  Tr A + b \  Tr \bar A \ ,
\label{inv}
\end{equation}
where $a$ and $b$ are constants roughly independent 
of $n$. On the other hand, it is not 
difficult to prove that the traces of the angular integrals appearing
in these relations 
are given by
\begin{equation}
Tr A = 1 \ ,  \qquad Tr \bar A = 2 n \ , \qquad
Tr C = 2 n \ ,  \qquad Tr \bar C = 4 {\cal I} \ .
\label{traces}
\end{equation}
As shown in Ref.\cite{HMS97}, ${\cal I} \le n^2$. In fact, from Table I we see that
${\cal I}$ is basically proportional to $n^2$. Therefore, replacing
Eq.(\ref{traces}) in Eq.(\ref{inv}) we obtain that $K^J$ should decrease as $n^2$ 
while $K^I$ goes only like $1/n$. This behavior of the inertia parameters has
important consequences in the multibaryon spectra. Namely, as the baryon number
increases low lying non-strange states are expected to have the lowest possible value 
of isospin. For strange multibaryons this is not necessarily the case due to the 
coupling of the isospin to the kaonic spin $T$.

The rotational energies for the non-strange multibaryons are given in Table VIII
while those for $S=-1$ states are given in Table IX and the corresponding to
zero-hypercharge states in Table X. In all the cases we have included in the tables
the lowest lying state and the first two excited states for each channel.
Some general observations can be made. Due to the overall decrease of the inertia 
parameters the energy splittings become smaller as $B$ increase.  
We also note that the ordering of the $S=0$ states is the same for both sets of parameters. 
For the $S=-1$ states there is, however, one exception which corresponds
to the second 
excited multibaryon with $(B,S) = (6,-1)$. For Set A the second 
excited state is 
a $3^+$ while for Set B is a $2^+$. It should be noticed, however, that the third excited
states (not listed in Table IX) are precisely a $2^+$ for Set A and $3^+$ for Set B
and that the energy difference with the second
excited state is $1 MeV$ in both cases. 
For the $Y=0$ states the situation becomes more complicated as $B$ increases. This is due to 
the rather small energy splittings between the different states.
As a general trend we also note that the rotational energies are slightly
smaller for Set B. This can be traced back to the fact that the
moments of inertia are 
smaller for that set of parameters. 

As discussed above, for non-strange baryons the lowest lying states always have the lowest
possible value of isospin. The corresponding spins are then given by the lowest
value allowed by the symmetry constraints. As remarked in Ref.\cite{Irw98} these values
turn out to be consistent with those known for light nuclei with the exception of
the odd values $B=5,7,9$. It should be stressed that at this point there is no obvious
way to identify these rather compact multiskyrmion configurations with normal nuclei.
Indeed, even for the $B=2$ case it is not clear to which extent the deuteron 
wave function in the Skyrme model is represented by the torus configuration. Some analysis
in terms of classical periodic orbits indicate that the two skyrmions spend most of their
time at large separation and only a short time near the torus\cite{Sny92}. 
As the strangeness increases (in absolute value) the quantum numbers of the low-lying
states become less obvious. This is a consequence of the interplay between the different
terms in the corresponding collective Hamiltonian for non-zero values of $T$. In fact, the quantum
numbers of the $Y=0$ states listed in Table X could be determined only after the calculation
of the energies of a rather large set of allowed states.  

We discuss now the issue of the stability of the $Y=I=0$ states that we generically call
multilambda states. The possible stability of a tetralambda state was first suggested in 
Ref.\cite{IKS88}. A similar conclusion was reached in Ref.\cite{SS98} where the existence
of a stable heptalambda was also proposed. As already mentioned in the Introduction, in that
work non-adiabatic corrections were neglected. We are now in position to
check whether these effects do or do not affect the stability of these states. From 
Table X we observe that for Set B the g.s. $Y=0$ tetrabaryon is indeed a
tetralambda state. 
This differs from the situation for Set A where the tetralambda is the
first excited state. In any case, this does not affect the rotational contribution to 
the ${4\Lambda}-{2\Lambda}$ mass difference. Using the energies given in Table X together
with the values given in Table V for the parameters of $H^{coll}_{B=2}$ (see Eq.(\ref{hdos})) 
we find that the rotational corrections decrease the binding by $36 \ MeV$ for Set A and
by $26 \ MeV$ for Set B. These values are significantly smaller than the binding energy
$\approx 176 \ MeV$ obtained for both sets of parameters in the adiabatic 
approximation\cite{SS98}. Thus, although the rotational corrections tend
to decrease the binding, the tetralambda still turns out to be bound within the present
approach. For the heptalambda we consider first its stability with respect to the decay 
into $3\Lambda + 4\Lambda$. The non-adiabatic value of the corresponding binding energy is
$-177 \ MeV$ \cite{SS98}. It should be noticed that the rotational energy of the
zero-isospin $(B, S)=(7,-7)$ state does not appear in Table X. In fact, the lowest lying
of such states has $E_{rot} = 104 MeV$ for Set A and $E_{rot} = 61 MeV$ for Set B. That is, 
it shows up as an excited state with higher energy. Nevertheless, taking into account the rather large
rotational energies of the $Y=I=0$ states with $B=3$ and $4$ it
happens that the binding energy
of the heptalambda is increased by $45 \ MeV$ for both sets of parameters. For the case of
the heptalambda ionization energy one can verify that the values given in Ref.\cite{SS98}
remain basically unaffected by the rotational corrections. For this purpose one has to
use the values of the rotational energies of the lowest $Y=I=0$ with $B=6$. Such
values (which are not listed in Table X) are $87 \ MeV$ for Set A and $52 \ MeV$ for Set B.

\section{Conclusions}

In this work we have studied the non-adiabatic corrections to the masses of the multibaryons
within the bound state approach to the $SU(3)$ Skyrme model. To describe the multiskyrmion
backgrounds we have used ans\"atze based on rational maps. Such configurations are
known to provide a good approximation to the exact numerical ones, and lead to
a great simplification in the solution of the kaon eigenvalue equation. An important
property of these approximate configurations is that they have the same symmetries 
as the exact ones. Consequently, the collective Hamiltonians and wave functions determined 
in this work are valid also in that case. They have been obtained making extensive use of the
properties of the corresponding symmetry groups. In particular, we have shown how the 
Finkelstein-Rubinstein phases fix, in a unique way, the one dimensional irreducible 
representations in which each wave function should transform.  

Using two standard sets of parameters for the effective $SU(3)$ Skyrme action we have
calculated all the inertia parameters and hyperfine splitting constants for $B \leq 9$.
We have found that as a general trend the isospin moments of inertia increase as $n^2$
while the spin ones as $n$, where $B=n$. Thus, the low lying non-strange multibaryons 
have the lowest possible value of isospin. The situation is more complicated in the
case of strange particles for which there is a quite delicate interplay between the 
different terms contributing to the rotational energies.  

We have also estimated the non-adiabatic corrections to the tetralambda and heptalambda
binding energies given in Ref.\cite{SS98}. We found that these corrections are
relatively small and do not affect the stability of these particles. This statement
can be certainly extended to the recent studies on the stability of heavier
flavored multiskyrmions\cite{KW99}.

We finish with a comment on the Casimir corrections to the multibaryon masses. Although
these corrections are not expected to affect in any significant way the rotational energies obtained
in the present work they might play some role in the determination of the
multibaryon binding energies. Within the $SU(2)$ Skyrme model it has been shown\cite{MK91} that 
they are responsible for the reduction of the otherwise large $B=1$ soliton mass to a reasonable value 
when the empirical value of $f_\pi$ is used. Here, we have avoided the $B=1$ large mass problem by using 
the customary method of fitting $f_\pi$ to reproduce the nucleon mass\cite{ANW83}. 
A more consistent approach should certainly use the empirical $f_\pi$ and include
the Casimir corrections. In this respect, there have been recently some
efforts\cite{Wal98} to evaluate the corrections to the $B=1$ mass in the $SU(3)$ Skyrme 
model. Unfortunately, even in the $SU(2)$ sector, almost nothing is known for $B > 1$. 
This is, of course, a very difficult task. Already in the $SU(2)$ model, it requires the
knowledge of the pion excitation spectrum around the non-trivial multiskyrmion up to rather 
large values of angular momentum. Nevertheless, recent studies of the $SU(2)$ multiskyrmion
low lying vibrational spectra\cite{BBT97} could be considered as first steps in this direction.

\acknowledgements
J.P.G. wishes to thank the kind hospitality of Laboratorio TANDAR,
CNEA, where part of this work was done. NNS wants to acknowledge
very useful discussions with E. Burgos, F. Parisi and C.L. Schat.
This work was supported in 
part by a grant from Fundaci\'on Antorchas, Argentina, and 
the grants PICT 03-00000-00133 and PMT-PICT0079 from ANPCYT, Argentina.
The work of J.P.G. was supported by EC Grant ARG/B7-3011/94/27.

\appendix

\section*{}

In this Appendix we present the explicit expression of the rational maps used
in this work. They are
\bea
R_1 &=& z \ , \\
R_2 &=& z^2  \ , \\
R_3 &=& { i \sqrt{3} \ z^2 - 1 \over{ z ( z^2 - i \sqrt{3} ) } }  \ , \\
R_4 &=& { 1 + 2i \sqrt{3} \ z^2 + z^4 
          \over{ 1 - 2 i \sqrt{3} \ z^2 + z^4} } \ , \\
R_5 &=& \frac{z\left(z^4 - i b_5 z^2 - 
	a_5\right)}{a_5 z^4 + i b_5 z^2 - 1} \ , \\
R_6 &=& \frac{z^4 + i a_6 }{z^2 \left(i a_6  z^4 + 1\right)} \ , \\
R_7 &=& \frac{z^5 - a_7 }{z^2 \left( a_7  z^4 + 1\right)} \ , \\
R_8 &=& \frac{ z^6 - i a_8 } {z^2 \left( i a_8  z^6 - 1\right) } \ , \\
R_9 &=& \frac{ z^3 \left( - z^6 + 3 i \sqrt 3 z^4 + 
	9 z^2 + 5 i \sqrt 3 \right) +
        a_9 z \left( - i \sqrt 3 z^6 - z^4 + i \sqrt 3 z^2 + 1 \right) }
 	{5 i \sqrt 3 z^6 + 9 z^4 + 3 i \sqrt 3 z^2 - 1 + 
        a_9 z^2 \left( z^6 + i \sqrt 3 z^4 - z^2 - i \sqrt 3 \right)} \ .
\eea
The numerical values
of the real constants $a_i, b_i$ appearing in these expressions are
\begin{equation}
a_5 = 3.07 \ , \quad a_6 = 0.158 \ , \quad a_7 = 0.143  
\ , \quad a_8 = 0.137  \ , \quad a_9 = 1.98 \ , \quad
b_5 = 3.94 \ .
\end{equation}
The reader can check that in most cases our maps agree with those given 
in Ref.\cite{HMS97}. There are a few exceptions, 
however. For $B=7$ we have
choose a different orientation in the spin and 
isospin spaces in such a way that
one of the 5-fold axes coincides with the z-direction. In the case of $B=9$
we have selected the map for which the $T_d$ group operations are 
realized in exactly the same way in both spin and isospin spaces (namely,
$g = D_g$). This is not the case for the $B=9$ map given in Ref.\cite{HMS97}.

\newpage

\begin{center}

\begin{table}
\mediumtext

\caption{Values of the diagonal elements of the angular integrals appearing in
Eqs.(\ref{integrals}). Also listed in Table 1 are the values of ${\cal I}$}

\begin{center}
\begin{tabular}{cccccccc}
n  & ${\cal I}$  &  $A$   &$\bar A$&   $ B $ &$\bar B$& $C$    &$\bar
C$\\ 
\hline
1  &   1         &   1/3  &  2/3   &  -2/3   &  -4/3  &  2/3   &  4/3
\\ 
\hline
2  & $\pi + 8/3$ &$1-\pi/4$&  4/3  &   0     &   0    &$\pi-2$ & $2\pi$ \\
   &             &$1-\pi/4$&       &   0     &   0    &$\pi-2$ & $2\pi$ \\
   &             &$\pi/2-1$&       & $\pi-4$ & -16/3  &$8-2\pi$& 32/3
\\ 
\hline 
3  &  13.58      &   1/3  & 2      &  0.348  &   4    &  2     & 18.11
\\ 
\hline 
4  &  20.65      & 0.391 &  8/3   &      0  &   0    &  8/3   & 27.53 \\ 
   &             & 0.391 &        &         &        &        &        \\
   &             & 0.219 &        &         &        &        &
\\ 
\hline
5  &  35.75      & 0.280 & 10/3   &  -0.090  &-3.649 & 3.440 & 50.65 \\ 
   &             & 0.280 &        &  -0.090  &-3.649 & 3.440 & 50.65 \\ 
   &             & 0.440 &        &  -0.051  &-2.038 & 3.119 & 41.71
\\ 
\hline
6  &  50.76      & 0.356  &  4     &    0    &   0    & 4.137 & 71.82 \\
   &             & 0.356  &        &    0    &   0    & 4.137 & 71.82 \\
   &             & 0.285  &        &  0.089  & 6.378  & 3.725 & 59.40
\\ 
\hline
7  &  60.87     &  1/3    &  14/3  &  0      &   0     &  14/3  &
81.16 \\ 
\hline
8  &  85.63      & 0.312  &  16/3  &    0    &   0    &  5.171 & 105.89 \\
   &             & 0.312  &        &    0    &   0    &  5.171 & 105.89 \\
   &             & 0.376  &        & -0.044 & -8.91 &  5.658 & 130.74
\\ 
\hline
9  &  113.07     &  1/3   &  6     & -0.062 & -7.51  &    6   & 150.76 
\\
\end{tabular}
\end{center}
\end{table}

\begin{table}
\mediumtext

\caption{Symmetry group $G$, generators of the proper subgroup, 
their corresponding FR phases and
the parity operations for B=3--9. The directions 
of the 3-fold axes in $B=7$ are defined by the spherical angles
$(\phi_\alpha, \theta_\alpha)$ = $\left( \pi/5, \quad
\arccos\left[ \sqrt{ ( 5 + 2 \sqrt{ 5 } )/15} \right]\right)$
and 
$(\phi_\beta, \theta_\beta)$ = $\left( 3\pi/5, \quad 
\arccos\left[1/\sqrt{ 15 + 6 \sqrt{ 5 } }\right] \right)$.}

\label{symop}

\begin{center}
\begin{tabular}{cccrcrc}
 &  & 
\multicolumn{4}{c}{Generators of proper subgroup 
and FR phases} & Parity \\ \cline{3-6}    
$B$ & $G$    & $\{ g_1, D_{g_1} \}$         &  $\gamma_{g_1}$   &  
         $\{ g_2, D_{g_2} \}$         &  $\gamma_{g_2}$  & operation \\ \hline
 3 & $T_d$     & $\{ C_3^{xyz}, C_3^{xyz} \}$ &   1           &  
         $\{ C_2^{z}, C_2^{z} \}$     &   1           &
         $\{ C_4^{z}, C_4^{z} \}$              \\ 
 4 & $O_h$    & $\{ C_3^{xyz}, C_3^{z}   \}$ &   1           &  
       $\{ C_4^{z}, C_2^{x} \}$       &   1           &
       $\{ E , C_2^{z} \}$                    \\ 
 5 & $D_{2d}$    & $\{ C_2^{z}, C_2^{z} \}$     &   1           &  
         $\{ C_2^{x}, C_2^{x} \}$     &  -1           &
         $\{ C_4^{z}, C_4^{z} \}$              \\ 
 6 & $D_{4d}$    & $\{ C_2^{x}, C_2^{x} \}$     &  -1           &  
         $\{ C_2^{xy}, C_2^{y} \}$    &  -1           &
         $\{ C_8^{z}, \bar C_4^{z} \}$         \\ 
 7 & $I_h$    &  $\{ C_5^{z}, (C_5^{z})^3 \}$ &  1           &
     $\{ C_3^{\alpha}, (C_3^{\beta})^2 \}$ &  1           &
         $\{ E , E \}$                         \\ 
 8 & $D_{6d}$    & $\{ C_2^{x}, C_2^{x} \}$     &   1           &  
         $\{ C_6^{z}, \bar C_3^{z} \}$&   1           &
         $\{ C_{12}^{z}, \bar C_6^{z} \}$      \\ 
 9 & $T_d$    & $\{ C_3^{xyz}, C_3^{xyz} \}$ &   1           &  
         $\{ C_2^{z}, C_2^{z} \}$     &   1           &
         $\{ C_4^{z}, C_4^{z} \}$         \\
\end{tabular}
\end{center}

\end{table}

\begin{table}
\narrowtext

\caption{Allowed values of $I$ and $T$ for states with different strangeness for B=3--9.}
\label{isospin}

\begin{center}
\begin{tabular}{cccc}
$B$    & $S$   &                 $I$                             &
$T$           \\ 
\hline
3      &   0   &        1/2, 3/2,\ldots,9/2                         &   0            \\ 
       &  -1   &            0, 1,\ldots,4                            &  1/2           \\    
       &  -3   &            0, 1,\ldots,3                            &  3/2           \\  
\hline
4      &   0   &            0, 1,\ldots,6                            &   0            \\ 
       &  -1   &        1/2, 3/2,\ldots,11/2                        &   1/2          \\    
       &  -4   &            0, 1,\ldots,4                            &   2            \\ 
\hline
5      &   0   &        1/2, 3/2,\ldots,15/2                        &   0            \\ 
       &  -1   &            0, 1,\ldots,7                            &  1/2           \\    
       &  -5   &            0, 1,\ldots,5                            &  5/2           \\  
\hline
6      &   0   &            0, 1,\ldots,9                            &   0            \\ 
       &  -1   &        1/2, 3/2,\ldots,17/2                        &  1/2           \\    
       &  -6   &            0, 1,\ldots,6                            &   3            \\ 
\hline
7      &   0   &        1/2, 3/2,\ldots,21/2                        &   0            \\ 
       &  -1   &            0, 1,\ldots,10                           &  1/2           \\    
       &  -7   &            0, 1,\ldots,7                            &  7/2           \\  
\hline
8      &   0   &            0, 1,\ldots,12                           &   0            \\ 
       &  -1   &        1/2, 3/2,\ldots,23/2                        &  1/2           \\    
       &  -8   &            0, 1,\ldots,8                            &   4            \\ 
\hline
9      &   0   &        1/2, 3/2,\ldots,27/2                        &   0            \\ 
       &  -1   &            0, 1,\ldots,13                           &   1/2          \\    
       &  -9   &            0, 1,\ldots,9                            &   9/2          \\ 
\end{tabular}
\end{center}

\end{table}

\begin{table}
\mediumtext

\caption{Numerical values of the radial integrals appearing in the
expressions of moments of inertia and hyperfine splitting constants.}

\begin{center}
\begin{tabular}{ccccccccc}
    & \multicolumn{4}{c}{SET A} & \multicolumn{4}{c}{SET B} \\ 
    \cline{2-5} \cline{6-9}
 $B$  & $m_1 (fm)$   & $m_2 (fm)$   & $d_1$  &  $d_2 ~(\times~ .01)$  
    & $m_1 (fm)$   & $m_2 (fm)$   & $d_1$  &  $d_2 ~(\times~ .01)$  \\ \hline
 1  & 1.27  & 0.212   &  0.227 &  2.29  
    & 1.22  & 0.304   &  0.275 &  2.93 \\ 
 2  & 1.95  & 0.233   &  0.233 &  1.27  
    & 2.21  & 0.335   &  0.272 &  1.61 \\ 
 3  & 2.58  & 0.241   &  0.228 &  0.89  
    & 3.06  & 0.349   &  0.261 &  1.13 \\ 
 4  & 3.00  & 0.246   &  0.213 &  0.69  
    & 3.65  & 0.359   &  0.243 &  0.89 \\ 
 5  & 3.74  & 0.249   &  0.217 &  0.57
    & 4.53  & 0.363   &  0.244 &  0.73 \\ 
 6  & 4.32  & 0.251   &  0.214 &  0.48 
    & 5.23  & 0.366   &  0.239 &  0.62 \\ 
 7  & 4.65  & 0.253   &  0.205 &  0.43  
    & 5.67  & 0.371   &  0.229 &  0.55 \\ 
 8  & 5.39  & 0.254   &  0.208 &  0.38  
    & 6.51  & 0.371   &  0.230 &  0.49 \\ 
 9  & 6.09  & 0.255   &  0.209 &  0.34  
    & 7.30  & 0.371   &  0.230 &  0.44 \\
\end{tabular}
\end{center}

\end{table}

\begin{table}
\narrowtext

\caption{Parameters for B=2}

\begin{center}
\begin{tabular}{cccccc}
SET & $K_1^J ~(MeV)$ & $K_1^I ~(MeV)$ & $\Theta_3^I ~(fm)$ & $\bar c_1^I$ & $c_3^I$ \\ 
\hline
A   &   33.42 &  53.68  &   1.15       &   0.409      &  0.631  \\ 
B   &   27.63 &  45.20  &   1.40       &   0.306      &  0.562  \\       
\end{tabular}
\end{center}
\end{table}

\begin{table}
\mediumtext

\caption{Inertia parameters for B=3-9}

\begin{center}
\begin{tabular}{ccccccc}
	& \multicolumn{3}{c}{SET A} & \multicolumn{3}{c}{SET B} \\
\cline{2-4}
\cline{5-7}
$B$    	&$K^J ~(MeV)$
        &$K^I ~(MeV)$
        &$K^M ~(MeV)$
	&$K^J ~(MeV)$
        &$K^I ~(MeV)$
        &$K^M ~(MeV)$ \\ 
\hline
3         &   15.23    &   50.77   &    9.55  &  12.11  &   41.03  &    7.80   \\ 
\hline
4         &   8.66     &  39.70    &     0    &  6.72   &  30.98   &     0     \\
          &            &  39.70    &          &         &  30.98   &           \\             
          &            &  32.88    &          &         &  25.89   &           \\        
\hline
5         &   5.20  &  28.29  &  -1.17  &   4.03   &  22.30   &  -0.96   \\
          &   5.20  &  28.29  &  -1.17  &   4.03   &  22.30   &  -0.96   \\
          &   5.88  &  33.91  &  -0.89  &   4.57   &  26.47   &  -0.73   \\
\hline
6         &   3.67   &  26.12  &    0     &   2.84  &  20.45   &    0    \\      
          &   3.67   &  26.12  &    0     &   2.84  &  20.45   &    0    \\   
          &   4.25   &  24.48  &   1.23   &   3.31  &  19.28   &   1.04  \\
\hline  
7   	  &   3.09   &  23.06  &    0   &   2.38    &  17.90   &    0    \\
\hline 
8    	  &   2.39  &  19.48   &    0    &   1.85  &  15.28  &    0      \\
          &   2.39  &  19.48   &    0    &   1.85  &  15.28  &    0      \\  
          &   2.11  &  21.08   &  -0.61  &   1.63  &  16.50  &  -0.52    \\  
\hline            
9    	  &   1.78  &  17.75   &  -0.43  &   1.39  &  14.02  &  -0.36    \\ 
\end{tabular}
\end{center}

\end{table}

\begin{table}
\narrowtext

\caption{Hyperfine splitting constants for B=3-9}

\begin{center}
\begin{tabular}{ccccc}
	& \multicolumn{2}{c}{SET A} & \multicolumn{2}{c}{SET B} \\ 
\cline{2-3}
\cline{4-5}
$B$	&$\bar c^J$&$\bar c^I$&$\bar c^J$&$\bar c^I$\\
\hline
3    &  -0.62    &   0.55     &  -0.64   &   0.48     \\  
\hline
4    &   0    &   0.55     &   0    &   0.48   \\               
     &        &   0.55     &        &   0.48   \\   
     &        &   0.46     &        &   0.37  \\
\hline
5    &   0.22  &   0.48   &   0.23   &   0.41   \\
     &   0.22  &   0.48   &   0.23   &   0.41   \\
     &   0.15  &   0.57   &   0.16   &   0.51   \\
\hline
6    &    0    &   0.53   &    0     &   0.46  \\
     &    0    &   0.53   &    0     &   0.46  \\  
     &  -0.28  &   0.49   &  -0.30   &   0.43  \\
\hline
7    &    0    &  0.53    &    0     &  0.46   \\
\hline 
8    &    0      &  0.51   &    0     &  0.45  \\
     &    0      &  0.51   &    0     &  0.45  \\   
     &  0.28     &  0.55   &  0.31    &  0.49  \\
\hline      
9    &  0.23     &  0.52   &  0.25    &  0.46  \\   
\end{tabular}
\end{center}
\end{table}

\begin{table}
\caption{Quantum numbers and rotational energies for $S=0$ states}
\begin{center}
\begin{tabular}{ccccccccc}
  &\multicolumn{4}{c}{SET A} & \multicolumn{4}{c} {SET B}  \\ 
\cline{2-5}
\cline{6-9}
$B$    & $J^P$    &   $I$   &  $N$  &$E_{rot} [MeV]$&
         $J^P$    &   $I$   &  $N$  &$E_{rot} [MeV]$ \\  \hline
 3     &${1/2}^+$ & ${1/2}$ &   1/2   &  64  &
        ${1/2}^+$ & ${1/2}$ &   1/2   &  52 \\
       &${5/2}^-$ & ${1/2}$ &   1/2   & 147  &
        ${5/2}^-$ & ${1/2}$ &   1/2   & 117 \\
       &${3/2}^-$ & ${3/2}$ &   3/2   & 205  &
        ${3/2}^-$ & ${3/2}$ &   3/2   & 164 \\
\hline
 4     &$ 0^+   $ &    0    &   0   &   0  &
        $ 0^+   $ &    0    &   0   &   0 \\
       &$ 4^+   $ &    0    &   0   & 173  &
        $ 4^+   $ &    0    &   0   & 134 \\
       &$ 0^+   $ &    2    &   2   & 238  &
        $ 0^+   $ &    2    &   2   & 186 \\
\hline
 5     &${1/2}^+$ & ${1/2}$ &   1/2   &  28  &
        ${1/2}^+$ & ${1/2}$ &   1/2   &  22 \\
       &${3/2}^+$ & ${1/2}$ &   1/2   &  40  &
        ${3/2}^+$ & ${1/2}$ &   1/2   &  31 \\
       &${3/2}^-$ & ${1/2}$ &   1/2   &  44  &
        ${3/2}^-$ & ${1/2}$ &   1/2   &  34 \\
\hline
 6     &$ 1^+   $ &    0    &   0   &   7  &
        $ 1^+   $ &    0    &   0   &   6 \\
       &$ 3^+   $ &    0    &   0   &  44  &
        $ 3^+   $ &    0    &   0   &  34 \\
       &$ 0^+   $ &    1    &   1   &  52  &
        $ 0^+   $ &    1    &   1   &  41 \\
\hline
 7     &${7/2}^+$ &${1/2}$&   1/2   &  66  &
        ${7/2}^+$ &${1/2}$&   1/2   &  51 \\
       &${3/2}^+$ &${3/2}$&   3/2   &  98  &
        ${3/2}^+$ &${3/2}$&   3/2   &  76 \\
       &${9/2}^+$ &${3/2}$&   3/2   & 163  &
        ${9/2}^+$ &${3/2}$&   3/2   & 126 \\
\hline
 8     &$  0^+  $ &   0   &   0   &   0  &
        $  0^+  $ &   0   &   0   &   0 \\
       &$  2^+  $ &   0   &   0   &  14  &
        $  2^+  $ &   0   &   0   &  11 \\
       &$  1^+  $ &   1   &   1   &  44  &
        $  1^+  $ &   1   &   1   &  34 \\
\hline
 9     &$  1/2^+    $ &   1/2   &   1/2   &   14  &
        $  1/2^+    $ &   1/2   &   1/2   &   11 \\
       &$  5/2^-    $ &   1/2   &   1/2   &   30  &
        $  5/2^-    $ &   1/2   &   1/2   &   24 \\
       &$  7/2^-    $ &   1/2   &   1/2   &   39  &
        $  7/2^-    $ &   1/2   &   1/2   &   31 \\
\end{tabular}
\end{center}
\label{nonstr}
\end{table}

\begin{table}
\label{str}
\mediumtext
\caption{Quantum numbers and rotational energies for $S=-1$ states}
\begin{center}
\begin{tabular}{ccccccccc}
  &\multicolumn{4}{c}{SET A} & \multicolumn{4}{c} {SET B}  \\ 
\cline{2-5}
\cline{6-9}
$B$    & $J^P$    &   $I$   &  $N$  &$E_{rot} [MeV]$&
         $J^P$    &   $I$   &  $N$  &$E_{rot} [MeV]$ \\  \hline
 3     &${1/2}^+$ &    0    &${1/2}$&  38  &
        ${1/2}^+$ &    0    &${1/2}$&  29 \\
       &${1/2}^+$ &    1    &${1/2}$&  84  &
        ${1/2}^+$ &    1    &${1/2}$&  72 \\
       &${5/2}^-$ &    0    &${1/2}$& 122  &
        ${5/2}^-$ &    0    &${3/2}$&  95 \\
\hline
 4     &$ 0^+   $ & ${1/2}$ &${1/2}$&   6  &
        $ 0^+   $ & ${1/2}$ &${1/2}$&   7 \\
       &$ 4^+   $ & ${1/2}$ &${1/2}$& 180  &
        $ 4^+   $ & ${1/2}$ &${1/2}$& 141 \\
       &$ 0^+   $ & ${3/2}$ &${3/2}$& 191  &
        $ 0^+   $ & ${3/2}$ &${3/2}$& 144 \\
\hline
 5     &${1/2}^+$ &    0    &${1/2}$&  11  &
        ${1/2}^+$ &    0    &${1/2}$&   7 \\
       &${3/2}^+$ &    0    &${1/2}$&  23  &
        ${3/2}^+$ &    0    &${1/2}$&  17 \\
       &${3/2}^-$ &    0    &${1/2}$&  28  &
        ${3/2}^-$ &    0    &${1/2}$&  21 \\
\hline
 6     &$ 1^+   $ & ${1/2}$ &$  0  $&  12  &
        $ 1^+   $ & ${1/2}$ &$  0  $&  10 \\
       &$ 0^+   $ & ${1/2}$ &$  1  $&  32  &
        $ 0^+   $ & ${1/2}$ &$  1  $&  23 \\
       &$ 3^+   $ & ${1/2}$ &$  0  $&  49  &
        $ 2^+   $ & ${1/2}$ &$  1  $&  37 \\
\hline
 7     &${7/2}^+$ &   0   &${1/2}$&  54  &
        ${7/2}^+$ &   0   &${1/2}$&  40 \\
       &${3/2}^+$ &   1   &${1/2}$&  74  &
        ${3/2}^+$ &   1   &${1/2}$&  56 \\
       &${7/2}^+$ &   1   &${1/2}$&  75  &
        ${7/2}^+$ &   1   &${1/2}$&  60 \\
\hline
 8     &$  0^+  $ &${1/2}$&$  0  $&   3  &
        $  0^+  $ &${1/2}$&$  0  $&   3 \\
       &$  2^+  $ &${1/2}$&$  0  $&  18  &
        $  2^+  $ &${1/2}$&$  0  $&  14 \\
       &$  1^+  $ &${1/2}$&$  1  $&  28  &
        $  1^+  $ &${1/2}$&$  1  $&  21 \\
\hline
 9     &$  1/2^+    $ &   0     &    1/2  &   4  &
        $  1/2^+    $ &   0     &    1/2  &   3 \\
       &$  5/2^-    $ &   0     &    1/2  &  20  &
        $  5/2^-    $ &   0     &    1/2  &  15 \\
       &$  1/2^+    $ &   1     &    1/2  &  21  &
        $  1/2^+    $ &   1     &    1/2  &  18 \\
\end{tabular}
\end{center}
\end{table}

\begin{table}
\label{multistr}
\mediumtext
\caption{Quantum numbers and rotational energies for $Y=0$ states}
\begin{center}
\begin{tabular}{ccccccccc}
  &\multicolumn{4}{c}{SET A} & \multicolumn{4}{c} {SET B}  \\ 
\cline{2-5}
\cline{6-9}
$B$& $J^P$    &   $I$   &  $N$  &$ E_{rot} [MeV]$&
         $J^P$    &   $I$   &  $N$  &$E_{rot} [MeV]$ \\  \hline
 3     &${1/2}^+$ &    1    &${1/2}$&  50  &
        ${1/2}^+$ &    1    &${1/2}$&  45 \\
       &${3/2}^-$ &    0    &${3/2}$&  77  &
        ${3/2}^-$ &    0    &${3/2}$&  52 \\
       &${3/2}^-$ &    1    &${3/2}$& 123  &
        ${5/2}^+$ &    0    &${3/2}$&  89 \\
\hline
 4     &$ 0^+   $ &    2    &   2   &  51  &
        $ 0^+   $ &    0    &   2   &  43 \\
       &$ 0^+   $ &    0    &   0   &  72  &
        $ 0^+   $ &    2    &   0   &  54 \\
       &$ 0^+   $ &    1    &   1   & 109  &
        $ 0^+   $ &    1    &   2   &  77 \\
\hline
 5     &${1/2}^+$ &    1    &${3/2}$&  29  &
        ${1/2}^+$ &    1    &${3/2}$&  21 \\
       &${1/2}^-$ &    1    &${3/2}$&  32  &
        ${1/2}^-$ &    1    &${3/2}$&  23 \\
       &${1/2}^+$ &    2    &${1/2}$&  39  &
        ${3/2}^+$ &    1    &${3/2}$&  30 \\
\hline
 6     &$ 0^+   $ &    2    &   1   &  24  &
        $ 0^-   $ &    1    &   2   &  16 \\
       &$ 0^-   $ &    1    &   2   &  26  &
        $ 1^-   $ &    1    &   2   &  22 \\
       &$ 1^-   $ &    1    &   2   &  33  &
        $ 1^+   $ &    1    &   2   &  23 \\
\hline
 7     &${3/2}^+$ &   2   &${7/2}$&  32  &
        ${3/2}^+$ &   2   &${7/2}$&  28 \\
       &${5/2}^+$ &   1   &${7/2}$&  65  &
        ${5/2}^+$ &   1   &${7/2}$&  42 \\
       &${7/2}^+$ &   1   &${7/2}$&  87  &
        ${7/2}^+$ &   1   &${7/2}$&  59 \\
\hline
 8     &$  0^+  $ &   2   &   2   &  19  &
        $  0^+  $ &   2   &   2   &  16 \\
       &$  2^+  $ &   2   &   2   &  31  &
        $  2^+  $ &   2   &   2   &  24 \\
       &$  2^+  $ &   2   &   2   &  33  &
        $  2^+  $ &   2   &   2   &  27 \\
\hline
 9     &$  1/2^-    $ &   2     &    5/2  &  25  &
        $  1/2^-    $ &   2     &    5/2  &  18 \\
       &$  3/2^-    $ &   2     &    5/2  &  29  &
        $  3/2^-    $ &   2     &    5/2  &  21 \\
       &$  3/2^+    $ &   3     &    3/2  &  31  &
        $  3/2^+    $ &   2     &    5/2  &  24 \\
\end{tabular}
\end{center}
\end{table}

\end{center}

\end{document}